\documentstyle[12pt]{article}
\begin{document}

\begin{center}
{\bf MECHANISMS OF THE REACTION $\pi^-p\rightarrow a^0_0
(980)n\rightarrow\pi^0\eta n$ \\ AT HIGH ENERGIES\\ [0.5cm]
N.N.~Achasov and G.N.~Shestakov}\\ [0.5cm]
{\it Laboratory of Theoretical Physics,
S.L. Sobolev Institute for Mathematics,\\
630090, Novosibirsk 90, Russia}\\ [0.5cm]
{\bf Abstract} \end{center}

Over the last twenty years, it was assumed that the reaction
$\pi^-p\to a^0_0(980)n\to\pi^0\eta n$ at high energies is dominated by the $b_1
$ Regge pole exchange. We show that the new Brookhaven and Serpukhov data on
the $t$ distributions for this reaction can be explained within the framework
of the Regge pole model only if the $\rho_2$ Regge pole exchange conspiring
with its daughter plays a crucial role. The tentative estimates of the absolute
$\pi^-p\to a^0_0(980)n\to\pi^0\eta n$ reaction cross section are obtained.
The contributions of the one-pion exchange mechanism and the Regge cuts are
also estimated.\\[0.5cm]

{\bf 1.}
In the $q\bar q$-model ($q$ is a light quark), every rotational excitation
with the orbital angular moment $L$ consists of four nonent: states $^{2S+1}L_J
=$ $^3L_{L-1}$, $^3L_L$, and $^3L_{L+1}$ with charge-parity $C=(-1)^{L+1}$ and
state $^1L_L$ with $C=(-1)^L$. But, so far there is a white spot in the
lower-lying family with $L=2$ [1].
The nonstrange members of the $^3D_2$ nonet with
quantum numbers $I^G(J^{PC})=1^+(2^{--})$ and $0^-(2^{--})$, i.e. the $\rho_2$,
$\omega_2$, and $\phi_2$ mesons (the masses of which are expected near 1.7 and
1.9 GeV [2,3]), are not yet identified as peaks in corresponding multibody
mass spectra [1]. However, the mass distributions are not unique keepers of
information on the resonances. The resonance spectrum is also reflected in the
Regge behavior of the reaction cross sections at high energies.
At present detailed investigations of the reaction $\pi^-p\to a^0_0(980)n$
at $P^{\pi^-}_{lab}=18$ and 38 GeV/c are carried out, respectively, at
Brookhaven [4] and Serpukhov [5]. This reaction involves only unnatural parity
exchanges in the $t$-channel. This fact is exclusive. Just it permits us to
conclude that the observed forward peak in the differential cross section of
$\pi^-p\to a^0_0(980)n$ is dominated by the Regge pole with quantum numbers
of the ``lost'' $\rho_2$ meson [6].

{\bf 2.} The $s$-channel helicity amplitudes for the reaction $\pi^-(q_1)p(p_1)
\to a^0_0(980,q_2)n(p_2)$ can be written as
\begin{equation} M_{\lambda_n\lambda_p}=\bar u_{\lambda_n}(p_2)\gamma_5[-A
-\gamma^\mu(q_1+q_2)_\mu B/2\,]u_{\lambda_p}(p_1)\ ,\end{equation}
where $\lambda_p(=+,-)$ and $\lambda_n(=+,-)$ are the proton
and neutron helicities, $A$ and $B$ are the invariant amplitudes depending on
$s=(p_1+q_1)^2$ and $t=(q_1-q_2)^2$ and free from kinematical singularities.
In the c.m. system, $d\sigma/dt=(|M_{++}|^2+|M_{+-}|^2)/64\pi s|\vec q_1|^2$
and, at fixed $t<-1$ GeV$^2$ and $s\gg m^2_N$, the helicity amplitudes
have the form [6]
\begin{equation} M_{++}\approx-sB\ , \qquad  M_{+-}\approx\sqrt{t_{min}-t}\,A
\ .\end{equation}
As it is well known, the $t$-channel helicity amplitudes
with definite parity and free from kinematical singularities are most suitable
for Reggeization [7]. These amplitudes for the reaction $\pi^-a^0_0\to\bar pn$
have the form \begin{equation} G_{++}=tA+m_N(m^2_{a_0}-m^2_\pi)B\ ,
\qquad G_{+-}=B\ .\end{equation} In the Regge region,
$G_{++}$ and $G_{+-}$ behave like $s^\alpha$ and $s^{\alpha-1}$ [6].
It is easy to show that the amplitude $G_{++}$ has to contain the Regge pole
exchange with $I=1$, $G=+1$, signature $\tau=-1$, and``naturality''
$\tau P=-1$. The high-lying Regge trajectory with such quantum numbers is the
$b_1$ trajectory (the well-known $b_1(1235)$ meson is its lower-lying
representative). The second independent amplitude $G_{+-}$ has to contain the
Regge pole exchanges with $I=1$, $G=+1$, $\tau=+1$, and $\tau P=-1$ and here
the $\rho_2$ Regge trajectory is a leading one.
In the physical region of the $s$-channel, the $b_1$ and $\rho_2$ Regge pole
contributions can be written as $G_{++}^{b_1}=\beta_{b_1}(t)(s/s_0
)^{\alpha_{b_1}(t)}ie^{-i\pi\alpha_{b_1}(t)/2}$\  and $G_{+-}^{\rho_2}=\beta_{
\rho_2}(t)(s/s_0)^{\alpha_{\rho_2}(t)-1}e^{-i\pi\alpha_{\rho_2}(t)/2}$,\ where
$\beta(t)$, $\alpha(t)$, and complex factors are residues, trajectories,
and signature factors of the corresponding Regge poles, and $s_0=1$ GeV$^2$.
As it is now evident from Eqs. (2) and (3),
the leading contributions to the $s$-channel helicity amplitudes $M_{++}$
and $M_{+-}$ at high energies are caused by the $\rho_2$ and $b_1$ Regge
exchanges, respectively.

Expressing the invariant amplitudes $A$ and $B$ in terms of $G_{++}$
and $G_{+-}$, \begin{equation} A=[G_{++}-m_N(m^2_{a_0}-m^2_\pi)G_{+-}]/t\ ,
\qquad B=G_{+-}\ , \end{equation} we find the $1/t$ singular behavior for the
invariant amplitude $A$. To avoid this singularity, one has to impose a further
constraint equation between the $t$-channel helicity amplitudes, namely: at
$t=0$, $G_{++}=m_N(m^2_{a_0}-m^2_\pi)G_{+-}$. It is of interest to satisfy the
analyticity of $A$ at $t=0$ assuming the different Regge pole models.

If we have the $b_1$ and $\rho_2$ exchanges only and assume that the amplitudes
$G^{b_1}_{++}$ and $G^{\rho_2}_{+-}$ do not vanish as $t\to0$, we obtain two
relations, $\alpha_{b_1}(0)=\alpha_{\rho_2}(0)-1$, and
$\beta_{b_1}(0)=m_N\left(m^2_{a_0}-m^2_\pi\right)\beta_{\rho_2}(0)$,
the first of which is rather silly because, at the usual values of
$\alpha_{b_1}(0)\approx-(0.05\div0.3)$, it requires $\alpha_{\rho_2}(0)
\approx0.95\div0.7$.
For the $\rho_2$ trajectory heaving unnatural parity, this is evidently ruled
out. Of course, in order for the amplitude $A$ to be regular for
$t\to0$, one can accept that the amplitudes $G^{b_1}_{++}$ and
$G^{\rho_2}_{+-}$ are separately proportional to $t$. But, in this case, a dip
in $d\sigma/dt$ in the forward direction is predicted by Eqs. (2) and (3).
On the contrary, the Brookhaven and Serpukhov experiments [4,5] show a clear
forward peak. This means that the amplitude $M_{++}$ with
quantum numbers of the $\rho_2$ exchange does not vanish as $t\to0$.
In the framework of the Regge pole model, this can be attended only in the
case of a conspiracy of the $\rho_2$ Regge trajectory with its daughter one
($d$), which has to have quantum numbers of the $b_1$ exchange. Let us
write down the contribution of such a daughter trajectory near $t=0$ in the
form $G_{++}^d=\beta_d(t)(s/s_0)^{\alpha_d(t)}ie^{-i\pi\alpha_d(t)/2}$.
Then, the amplitude $A$ should be regular at $t=0$ if the following relations
for the $\rho_2$, $d$, and $b_1$ exchanges are valid:
$\alpha_d(0)=\alpha_{\rho_2}(0)-1$,\  $\beta_d(0)=m_N(m^2_{a_0}-m^2_\pi)\beta_
{\rho_2}(0)$,\  $\beta_{\rho_2}(0)\not=0$, and $\beta_{b_1}(t)\sim t$. Thereby
the amplitude $M_{++}$ dominated by the $\rho_2$ exchange dos not vanish
as $t\to0$ and can lead to the observed forward peak. Note that asymptotically
(at large $s$) the daughter contribution and nonasymptotic contribution of the
$\rho_2$ trajectory (which behave as $\sim s^{\alpha_{\rho_2}-1}$) to the
amplitude $A$ and consequently to the amplitude $M_{+-}$ can all be neglected.

Thus, in the model with the $b_1$ and conspiring $\rho_2$ Regge poles, the
data on the $t$ distributions at fixed $s$
can be fitted to the form \begin{equation} dN/dt=C_1e^{\Lambda_1t}+(t_{min}-t)
\,C_2e^{\Lambda_2t}\ ,\end{equation} where the first term corresponds
to the contribution of the amplitude $M_{++}$ with the conspiring
$\rho_2$ exchange and the second term corresponds to the contribution of the
amplitude $M_{+-}$ with the $b_1$ exchange. The Brookhaven data
for $-t_{min}<-t<0.6$ GeV$^2$ [4] are fitted
by Eq. (5) equally well both with and without the $b_1$ exchange.
The fit with $C_2\equiv0$ in Eq. (5) gives $C_1=129$ events/GeV$^2$,
$\Lambda_1=4.7$ GeV$^{-2}$, and $\chi^2/$n.d.f.$\approx15.9/22$.
It is shown in Fig. 1 by the solid curve. The fit to these data to
the full form in Eq. (5) gives
$C_1=131$ events/GeV$^2$, $\Lambda_1=7.6$ GeV$^{-2}$, $C_2=340$ events/GeV$^4$,
$\Lambda_2=5.8$ GeV$^{-2}$, and $\chi^2/$n.d.f.$\approx15.9/20$.
The corresponding curve is practically the same as the solid one
in Fig. 1. The dashed and dotted curves in Fig. 1 show the $\rho_2$
and $b_1$ contributions separately. The latter yields approximately 34\% of the
integrated cross section.

In order to have an idea of the absolute value of the $\pi^-p\to a^0_0(980)n\to
\pi^0\eta n$ cross section, it has been estimated by us in Ref.
[6] at $P^{\pi^-}_{lab}=18$ GeV/c using the available data on the reaction $\pi
^-p\rightarrow a^0_2(1320)n$ and the Brookhaven data [4] on the $\pi^0\eta$
mass spectrum in $\pi^-p\rightarrow\pi^0\eta n$. Our tentative estimate is:
$\sigma(\pi^-p\rightarrow a^0_0(980)n\rightarrow\pi^0\eta n)\approx 200$ nb
and $[d\sigma/dt(\pi^-p\rightarrow a^0_0(980)n\rightarrow\pi^0\eta n)]_
{t\approx0}\approx940$ nb/GeV$^2$. Note that $(6-15)$\% of the integrated cross
section can be caused by the one-pion exchange mechanism which arises owing to
the $f_0(980)-a^0_0(980)$ mixing violating isotopic invariance [6]. Also
the $a_2\pi$, $\rho b_1$, $\omega a_1$, $\rho\rho$P, and $a_2a_2$P Regge cut
contributions (where P is the Pomeron) to the amplitude $M_{++}$ have been
estimated in Ref. [6] and the conclusion has been made that they are small.

Note that the early estimate of the $a_0(980)$ production [8] assumed the
$b_1$ exchange dominance and consequently a dip in $d\sigma/dt$ in the forward
direction.

{\bf 3.} Let us note two practical experimental problems. $i)$ It is
important to determine rather accurately the parameters of
the simplest Regge pole model. To do so it is needed the good data on
$d\sigma/dt(\pi^-p\rightarrow a^0_0(980)n)$ at several appreciably different
energies. First of all, we have in mind the energies of the $\pi^-$
beams at Serpukhov, Brookhaven, and KEK.
$ii)$ It would be very interesting to find some signs of the $\rho_2$ state and
its daughter state having the $b_1$ meson quantum numbers, for example,
in the $a_0\pi$, $\omega\pi$, and $A_2\pi$ mass spectra around 1.7 GeV.

\begin{flushleft} {\large \bf Figure caption} \end{flushleft} 
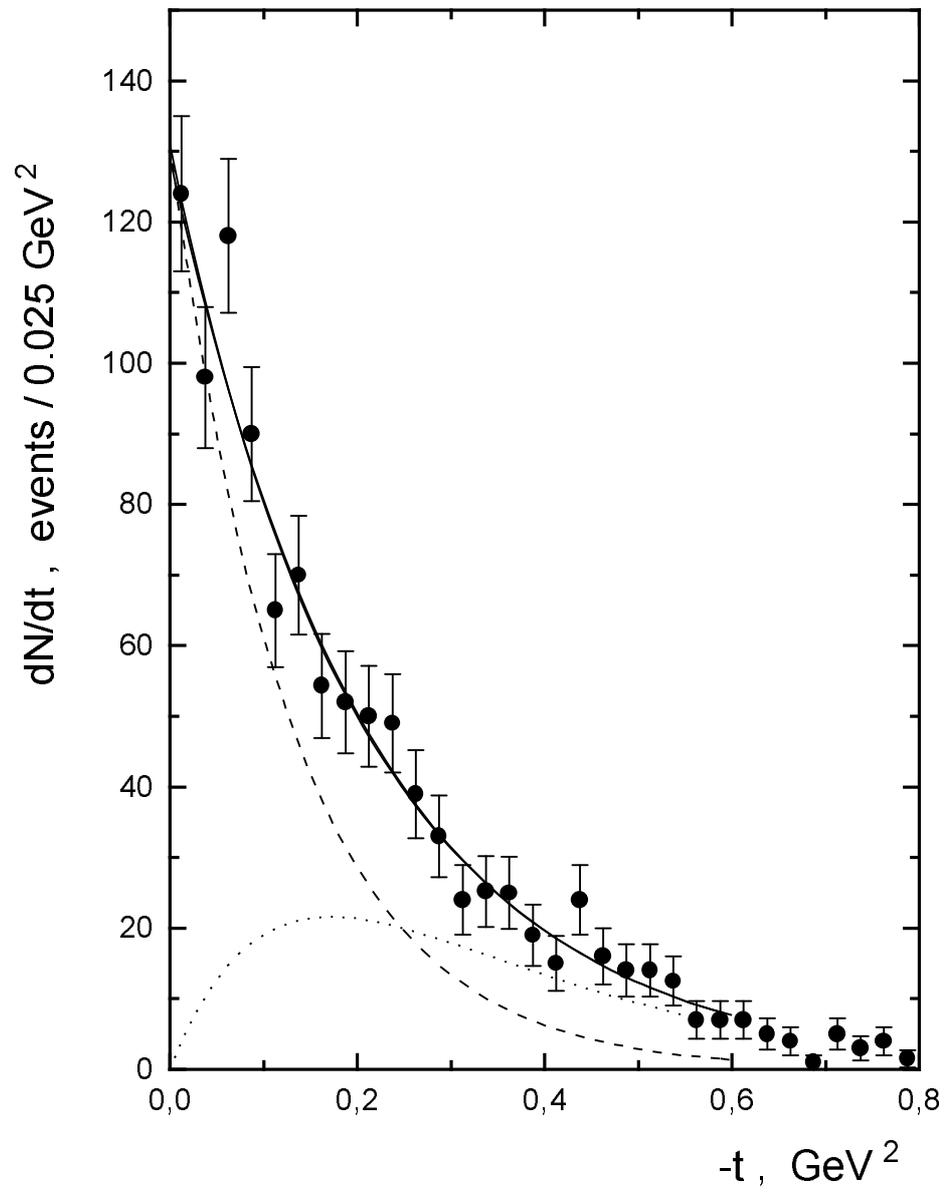
{\bf Fig. 1.} \ The $t$ distribution for the reaction $\pi^-p\rightarrow a^0_0(
980)n\rightarrow\pi^0\eta n\rightarrow4\gamma n$ at 18 GeV/c measured at
Brookhaven [4]. The fits are described in the text.

\end{document}